\documentclass[a4paper]{article}

\usepackage{INTERSPEECH2022}

\title{Training speaker recognition systems with limited data}
\name{Nik Vaessen$^1$, David A. van Leeuwen$^1$}
\address{
  $^1$Institute for Computing and Information Sciences, Radboud University }
\email{nvaessen@science.ru.nl, dvanleeuwen@science.ru.nl}

\usepackage{microtype}
\usepackage{graphicx}
\usepackage{subfigure}
\usepackage{booktabs} 
\usepackage{cite} 

\usepackage{hyperref}

\usepackage{amsmath}
\usepackage{amssymb}
\usepackage{mathtools}
\usepackage{amsthm}
\usepackage{booktabs}
\usepackage{float}
\usepackage{placeins}

\usepackage[capitalize,noabbrev]{cleveref}
\usepackage{blindtext}

\crefformat{section}{\S#2#1#3} 
\crefformat{subsection}{\S#2#1#3}
\crefformat{subsubsection}{\S#2#1#3}

\def\math#1{\ifmmode#1\else$#1$\fi}

\let\underscore=\_
\def\_{\checkmath_\underscaore}
\def\checkmath#1#2{\ifmmode\def\next##1{#1{\rm##1}}\else\let\next=#2\fi\next}

\newcount\exponent
\def\e{\afterassignment\ee\exponent=}
\def\ee{\math{10^{\the\exponent}}}

\def\negskip{\vskip-\baselineskip}

\makeatletter

\renewcommand{\section}{\@startsection
  {section} 
  {1} 
  {0} 
  {-0.4\bigskipamount} 
  {0.4\bigskipamount} 
  {}} 

\renewcommand{\subsection}{\@startsection
  {subsection} 
  {2} 
  {} 
  {-0.8\medskipamount} 
  {0.8\medskipamount} 
  {} 
}

\renewcommand{\subsubsection}{\@startsection
  {subsubsection} 
  {3} 
  {} 
  {-\smallskipamount} 
  {\smallskipamount} 
  {} 
}

\renewenvironment{thebibliography}[1]{%
  \begin{oldthebibliography}{#1}%
    \setlength{\parskip}{0.1ex}%
    \setlength{\itemsep}{0.1ex}%
  }%
  {%
  \end{oldthebibliography}%
}

\begin{document}

\maketitle
\begin{abstract}

This work considers training neural networks for speaker recognition with a much smaller dataset size compared to contemporary work. We artificially restrict the amount of data by proposing three subsets of the popular VoxCeleb2 dataset. These subsets are restricted to 50\,k audio files (versus over 1\,M files available), and vary on the axis of number of speakers and session variability. We train three speaker recognition systems on these subsets; the X-vector, ECAPA-TDNN, and wav2vec2 network architectures. We show that the self-supervised, pre-trained weights of wav2vec2 substantially improve performance when training data is limited. Code and data subsets are available at \url{https://github.com/nikvaessen/w2v2-speaker-few-samples}.
\end{abstract}


\section{Introduction}

Recently, the wav2vec2 framework~\cite{Baevski2020Wav2vec2} proposed a self-supervised pre-training, and consecutive fine-tuning approach for automatic speech recognition with a transformer network. Such a procedure has become the de facto standard in NLP with models like BERT \cite{devlin2018bert}. One of the benefits of pre-training is the possibility to use large, unlabeled, and thus relatively inexpensive datasets. Another benefit is that these pre-trained networks are flexible, and can be fine-tuned to a variety of related tasks. This has been shown to be the case for wav2vec2 as well, which, while originally designed for speech recognition~\cite{Baevski2020Wav2vec2}, has also been used for tasks like speaker recognition~\cite{Fan21ExplWav2vec2,Vaessen2021Wav2vec2FineTuneSpRec,evain2021Lebenchmark} and emotion recognition~\cite{pepino2021wav2vec2emotion,yuan2021LhoneEmotion,evain2021Lebenchmark}. One property of fine-tuning a pre-trained network is that it requires less labeled data than training from scratch. For example, the authors of wav2vec2 pre-train on 53\,k hours of unlabeled speech data, fine-tune on 10 minutes of labeled speech data, and achieve a word-error-rate of 4.8\% on the clean test set of LibriSpeech~\cite{panayotov2015librispeech}. For comparison, in 2016 the DeepSpeech2 system~\cite{Amodei2016DeepSpeech2} achieved a 5.3\% word-error-rate with 3600 hours of labeled training data.  

In this work, we want to study the behaviour of wav2vec2 under similar low-resource data conditions, but for speaker recognition instead of speech recognition. We are interested in the following research questions:

\begin{enumerate}
    \item How well does the self-supervised, pre-trained wav2vec2 network perform when fine-tuned for speaker recognition with little labeled data?
    \item What is the most effective way to structure a limited training dataset? Is there a trade-off to be made between speaker variability and session variability?
\end{enumerate}

We hypothesize that the pre-trained wav2vec2 network will also be beneficial for the limited-data speaker recognition scenario, as it has learnt to model speech representations. This was shown useful as a basis for speech recognition, and it seems plausible that speaker recognition can benefit from representations of speech. Although we compare wav2vec2 against non-self-supervised neural networks designed specifically for speaker recognition~\cite{snyder2018x,desplanques20_interspeech}, we imagine that these (or similar) networks can also benefit from self-supervision. There has been work for speaker recognition on self-supervised learning~\cite{Cai2021iterative,Thiendpondt2021Voxsrc20}, and consecutive fine-tuning~\cite{Chen2021Wavlm}, but to the extent of our knowledge, not for common speaker recognition networks~\cite{snyder2018x,desplanques20_interspeech}. Also, note that a frequent solution to limited data is data augmentation~\cite{shorten2019dataaug}. In this work, we explicitly skip data augmentation in order to observe the effects of self-supervised weights. The second research question is focused on data collection. There might be scenarios, related to e.g., licensing, or the domain, where one needs to construct a dataset for fine-tuning. In this scenario, we hypothesize that maximizing the number of speakers in your dataset is paramount. 

\section{Related work}

Earlier work~\cite{poddar2018speaker,das14_interspeech,jayanna2009experimental} interprets limited data availability not in the size of the training dataset, but in the length of the utterances. However, since the advent of neural approaches for speaker recognition, it has become the norm to train with short audio segments, often between 0.5 and 3 seconds~\cite{snyder2018x,desplanques20_interspeech,lin2020wav2spk}. The contemporary field of few-shot learning~\cite{Wang2020FewShot} considers low resource scenarios where models need to adapt to new classes (``N-way'') with only a few samples (``K-shot''). 
In \cite{Li2020SpRecLimData}, the LibriSpeech dataset~\cite{panayotov2015librispeech} is used to study low resource conditions for speaker identification. They vary the total training data length per speaker between 20, 40 or 60 segments of 3 seconds, and show only minor degradation in test accuracies when using a prototypical loss~\cite{snell2017prototypical}, or their proposed adversarial few-shot learning-based speaker identification framework. 
In \cite{Wang2019CentroidMetrSpRec}, speaker verification is considered within the few-shot learning paradigm with a subset of VoxCeleb2~\cite{Chung18b} containing 71 train speakers and 30 test speakers. They compare a prototypical loss~\cite{snell2017prototypical} against a triplet loss~\cite{zhang17d_interspeech}, and train with 200 segments of 2 seconds. They show prototypical loss achieves better equal-error-rates than the triplet loss in this scenario.  



\section{Methodology} \label{sec:meth}

\begin{table*}[h!]
\centering
\caption{Statistics on the vox2 training split, as well as the three tiny subsets we created, that we used for training.}
\label{tab:data_stats}
\begin{tabular}{lrrrrrrrrrrrr}
\hline
                   & \multicolumn{1}{c}{}             & \multicolumn{1}{c}{}            & \multicolumn{1}{c}{}            & \multicolumn{1}{c}{}              & \multicolumn{1}{c}{} & \multicolumn{3}{c}{sessions per speaker} & \multicolumn{1}{c}{} & \multicolumn{3}{c}{utterances per sessions} \\ \cline{7-9} \cline{11-13} 
dataset            & \multicolumn{1}{c}{duration (h)} & \multicolumn{1}{c}{\# speakers} & \multicolumn{1}{c}{\# sessions} & \multicolumn{1}{c}{\# utterances} & \multicolumn{1}{c}{} & mean         & min         & max         &                      & mean          & min          & max          \\ \hline
vox2               & 2\,314                           & 5\,994                          & 136\,632                        & 1\,068\,871                       &                      & 22.8         & 4           & 89          &                      & 7.8           & 1            & 264          \\
tiny-few-speakers  & 113                              & 100                             & 5\,066                          & 49\,400                           &                      & 50.7         & 22          & 87          &                      & 9.8           & 1            & 264          \\
tiny-few-sessions  & 100                              & 5\,994                          & 6\,275                          & 47\,952                           &                      & 1.0          & 1           & 4           &                      & 7.6           & 1            & 8            \\
tiny-many-sessions & 97                               & 5\,994                          & 46\,813                         & 47\,952                           &                      & 7.8          & 4           & 8           &                      & 1.0           & 1            & 3            \\ \hline
\end{tabular}
\negskip
\end{table*}

\subsection{Subsetting VoxCeleb} \label{sec:meth-subset}

In order to experiment with smaller dataset size conditions, we artificially limit ourselves to a subset of data from the so-called development set of VoxCeleb2~\cite{Chung18b}, which we use as \emph{train} and \emph{validation} set.
This development set consists of nearly 6\,k speakers distributed over 1\,M speech utterances, with a mean length of $7.8$ seconds and a standard deviation of $5.2$ seconds. 
Each speaker has a number of associated video recordings (sessions), and from each recording one or more speech utterances are automatically extracted by using face tracking, face verification, and active speaker verification. 
This ensures each speech utterance is attributed to a single speaker, although some labeling noise is expected. Recordings were collected by \cite{Chung18b} using the top 100 results from a YouTube search. The search query included the name of the celebrity and the word “interview”. 
We limit each subset to 50\,k utterances, but vary the amount of speakers, the amount of sessions per speaker, or the amount of utterances per session. Throughout this work, we will refer to the following datasets. Statistics of the datasets are shown in Table \ref{tab:data_stats}. 

\textbf{Vox2} refers to our train split of the original dataset.
We also create a validation split for monitoring overfitting, and selecting the best checkpoint. For each speaker in the original VoxCeleb2 development set, we randomly move sessions to the validation dataset until less than $99\%$ of all utterances from the respective speaker are remaining. The validation set is created \textit{before} the tiny training subsets. To calculate a validation EER, we create a random trial list with 15\,k positive and negative (same-gender) trials. The validation set is equal for \textit{vox2}, \textit{tiny-few-sessions} and \textit{tiny-many-sessions}. For \textit{tiny-few-speakers}, we modify the validation set such that it only contains the 100 speakers in the subset, and a different random trial list, with 2\,k positive and 2\,k negative (same-gender) trials is created. 

\textbf{Tiny-few-speakers} is a subset with few speakers, but many sessions per speaker, and many utterances per session. We group the speakers from \textit{vox2} by gender, and sort descendingly by the amount of recordings available. We select the first 50 female and male speakers. 

\textbf{Tiny-few-sessions} is a subset with many speakers, few sessions per speaker, but relatively many utterances per session. This subset contains all speakers in \textit{vox2}. For each speaker, we sort their sessions descendingly by the number of utterances. We then select 8 utterances from each speaker. We start sampling from the session with the most utterances. If a session is exhausted, we continue with the next session according to the sorted collection. 

\textbf{Tiny-many-sessions} is a subset with many speakers, many sessions per speaker, but few utterances per session. Just as in \textit{tiny-few-sessions}, we select 8 utterances from all speakers in \textit{vox2}. However, the difference is that we sample only $1$ utterance per session. When a speaker has fewer than 8 sessions available, we cycle through the sorted collection, selecting only 1 utterance per session per cycle, until 8 utterances are selected.

\subsection{Speaker recognition networks}

We train three different speaker recognition models on the datasets in \cref{sec:meth-subset}. We use third-party library network implementations, but train and evaluate with our own PyTorch~\cite{Paszke2019Pytorch} code.

\subsubsection{X-vector}

The X-vector architecture~\cite{snyder2018x} is a popular neural network for speaker recognition and diarization, initially from the Kaldi framework~\cite{povey2011kaldi}. The X-vector network consists of 5 consecutive layers with  1-dimensional convolution, ReLU activation, and BatchNorm, followed by mean\&std pooling, and 2 FC layers. During training, a third FC layer is used for computing the classification loss. We extract the speaker embeddings from the first fully-connected layer. We use the implementation by SpeechBrain~\cite{speechbrain}, with the default settings, such that the speaker embeddings have a dimensionality of 1024.   

\subsubsection{ECAPA-TDNN}

The ECAPA-TDNN architecture~\cite{desplanques20_interspeech} is a more recent speaker recognition network that showed best performance in the VoxCeleb 2020 challenge~\cite{Nagrana2020VoxChallenge}. It builds on top of the X-vector paradigm by adding global context through network architecture modifications. First, it makes use of three consecutive res2blocks~\cite{gao2019res2net}, consisting of three 1-d convolutions, a squeeze-and-excitation layer~\cite{hu2018squeeze} and a skip connection~\cite{he2016deep}. They also aggregate the output of each res2block, before using channel-wise attentive statistical pooling to compute a fixed-size speaker embedding. We use the SpeechBrain~\cite{speechbrain} implementation, with the default settings, such that the speaker embeddings have a dimensionality of 128.

\subsubsection{Wav2vec2}

The wav2vec2 architecture applies self-supervised pre-training to speech data, and has been used for multiple speech-related tasks. We only fine-tune the network. There is BASE and LARGE variant, we only consider the BASE network. The architecture consists of 3 components: First, raw audio is processed by a 7-layer CNN with 1-d convolutions, LayerNorm, and GELU activation~\cite{hendrycks2016gaussian}. Secondly, a linear projection with a FC layer and an additive relative positional embedding with a 1-layer CNN is applied. Lastly, the hidden state sequence is processed by 12 transformer blocks. We use the Transformers~\cite{wolf-etal-2020-transformers} implementation, with self-supervised weights provided by Fairseq\footnote{We downloaded the weights from \url{https://huggingface.co/facebook/wav2vec2-base}}. The self-supervision was carried out (by Fairseq~\cite{Baevski2020Wav2vec2}) on the LibriSpeech~\cite{panayotov2015librispeech} dataset. For the speaker recognition task, the final hidden state sequence is mean-pooled into a fixed-size speaker embedding of dimensionality 768~\cite{Vaessen2021Wav2vec2FineTuneSpRec,Fan21ExplWav2vec2}. During training, a single FC layer is used for classification.

\section{Experiments}

The experiments consist of a hyperparamter search for the best learning rate for each network and dataset combination (\cref{sec:grid}), multiple runs with the best learning rate and varying amount of steps (\cref{sec:performance}), and an ablation study (\cref{sec:ablation}). 

\subsection{Training and evaluation protocol} \label{sec:trainprotocol}

We base the following training protocol on the ECAPA-TDNN~\cite{desplanques20_interspeech} and wav2vec2~\cite{Baevski2020Wav2vec2} articles. Each network is trained for $n\_{steps}$ steps with the Adam~\cite{adamkingma14} optimizer. We use a cyclic learning rate schedule~\cite{smith2017cyclical} with a degrading maximum learning rate according to the \emph{triangular2} policy. We always use 4 cycles, one cycle is therefore $n\_{steps}/4$ iterations. The minimum LR each cycle is $\e-8$. For the \textit{vox2} dataset we validate every 5\,k steps, for the tiny datasets we validate after each epoch. To create a batch, we randomly sample speech utterances, and select a random 2 second chunk from each utterance. We use a batch size of 100 chunks, matching the total batch size of 3.2\,M audio samples used in \cite{Baevski2020Wav2vec2}. For X-vector and ECAPA-TDNN the network input is a 80-dimensional MFCC with a window length of 25 ms and a 12.5 ms shift.
All three network are trained with angular additive margin softmax loss~\cite{Deng_2019_CVPR,liu19f_interspeech}. We use a margin of 0.2 and a scale of 30. We do not use any weight decay in order to reduce the search space. For X-vector and ECAPA-TDNN, we use SpecAugment~\cite{park19e_interspeech} with 5 to 10 masks of length 10 in the time axis, and 1 to 3 masks of length 4 in the channel axis. For wav2vec2, we use a LayerDrop~\cite{huang2016deep,fan2019reducing} of $10\%$ in the transformer layers, and a dropout of $10\%$ is applied after each fully-connected layer in the network. Wav2vec2 also applies masking before the relative positional embedding is added, similar to SpecAugment; $10\%$ of the channel dimensions are randomly masked, and $50\%$ of the time dimensions are randomly masked. We freeze the whole wav2vec2 network for the first 12.5\,k steps, except for the last fully-connected layer used for the speaker classification. We also freeze the feature extractor CNN of wav2vec2 for the whole training run ($n\_{steps}$ iterations). Training is conducted on a RTX 3090 GPU for wav2vec2, and an GTX 2080Ti GPU for X-vector and ECAPA-TDNN. All experiments are capped to 32\,GB RAM and 6 CPU cores. In total 209 days of GPU time was used for experiments.

We evaluate trials using a cosine score between speaker embeddings, without any other processing. We use the original VoxCeleb1~\cite{Nagrani17} test set (40 speakers, henceforth \textit{vox1-o}) as a development set, and the VoxCeleb1 hard test set ($1190$ speakers, henceforth \textit{vox1-h}) as the evaluation set. There is no overlap between VoxCeleb1~\cite{Nagrani17} and Voxceleb2~\cite{Chung18b}. There is an overlap between the development and evaluation set, but we verified that the results are not significantly different when the trials from overlapping speakers are removed.

\subsection{Learning rate search} \label{sec:grid}

We conduct a learning rate search for the maximum LR in the cyclic schedule. This is done for all three networks and all four datasets. We conduct this search in two phases. In the first phase we scan over a large magnitude: we consider $10^{-i},$ with $i \in \{2, 3, 4, 5, 6, 7\}$. Based on the development set, we select the LR $10^{-j}$ with the lowest EER. In the second phase, we scan around this LR: we consider $\{1.78,3.16,5.62\} \times \{10^{-j-1},10^{-j}\}$. After the second phase, the LR with the lowest EER is used for the remaining experiments. The random seed is kept constant across all training runs in the grid search, and thus every learning rate is attempted only once. Each run has $n\_{steps}= 50\,\text{k}$.

\begin{figure*}
    \centering
    \includegraphics[width=1\textwidth,keepaspectratio]{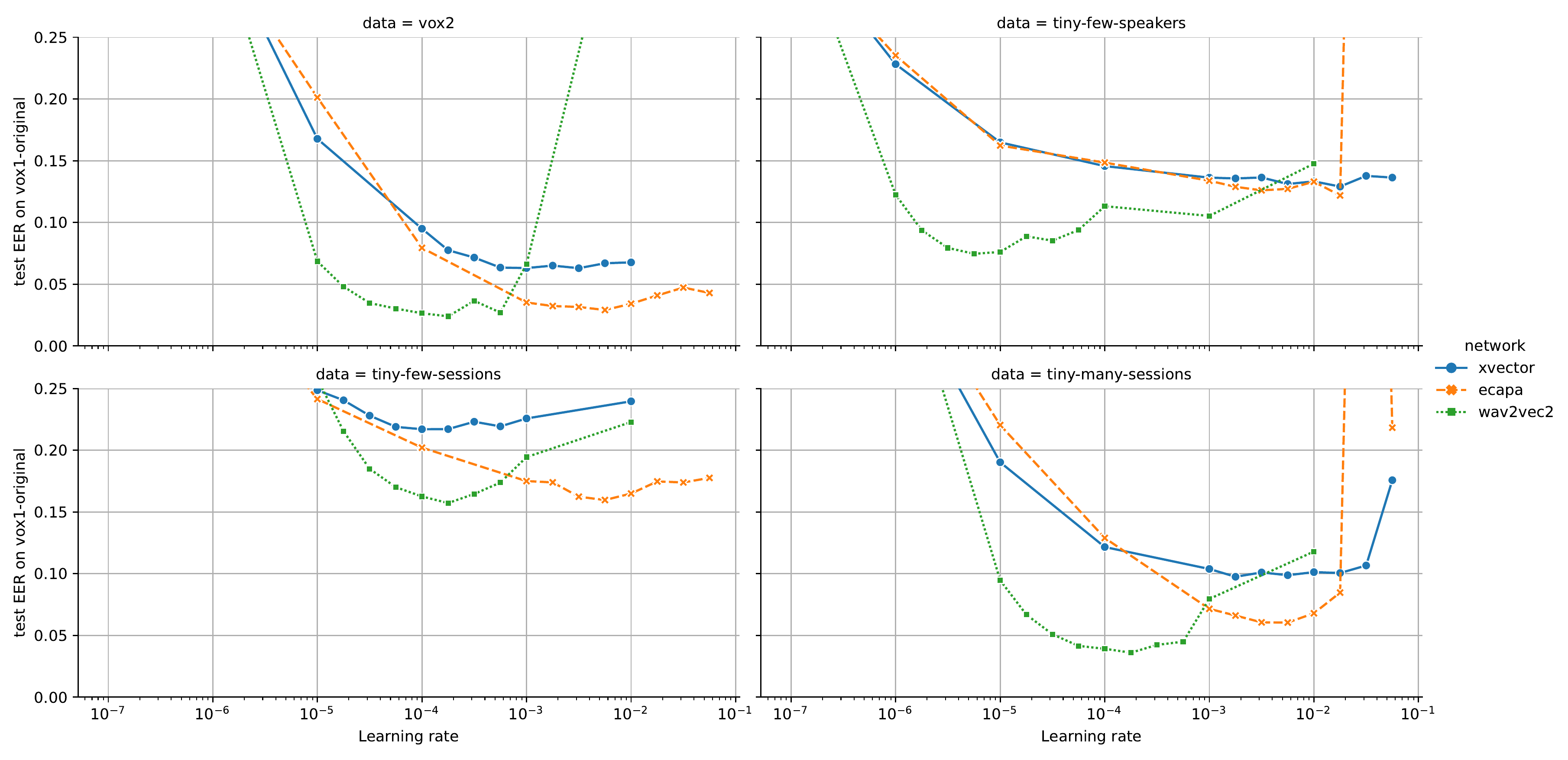}
    \caption{The results of the learning rate search. We plot the learning rates of phase 1 and phase 2 of the grid search on the x-axis, and the EER on the \textit{vox1-o} development set on the y-axis. Each dot represents a single experiment, no variation is measured.}
    \label{fig:grid_search}
    \negskip
\end{figure*}

The best LR, and the respective EER on the development set \textit{vox1-o}, are shown in Table \ref{tab:bestgrid}. We see that wav2vec2 performs best on all datasets. However, performance on \textit{tiny-few-sessions} seems poor for all three networks. 
In Figure \ref{fig:grid_search} we plot the learning rate against the EER. In general, we can see that wav2vec2 requires a lower learning rate. Moreover, for all three networks, the optimal learning rate is dependent on the dataset. 

\begin{table}[t]
\centering
\caption{The best-performing learning rate for each network and dataset combination, trained for 50\,k steps. The respective EER is measured on the \textit{vox1-o} development set. }
\label{tab:bestgrid}
\begin{tabular}{llccc}
\hline
data  &     & X-vector         & ECAPA            & wav2vec2             \\ \hline
vox2  & EER & 6.30\,\%           & 2.91\,\%           & 2.40\,\%               \\
      & LR  & $3.16\times\e-3$ & $5.62\times\e-3$ & $1.78\times\e-4$     \\ \hline
few   & EER & 12.91\,\%          & 12.19\,\%          & 7.46\,\%               \\
speak. & LR  & $1.78\times\e-2$ & $1.78\times\e-2$ & $5.62\times\e-6$ \\ \hline
few   & EER & 21.70\,\%          & 15.97\,\%          & 15.72\,\%              \\
sess.  & LR  & $1.00\times\e-4$ & $5.62\times\e-3$ & $1.78\times\e-4$     \\ \hline
many  & EER & 9.75\,\%           & 6.04\,\%           & 3.60\,\%               \\
sess.  & LR  & $1.78\times\e-3$ & $5.62\times\e-3$ & $1.78\times\e-4$     \\ \hline
\end{tabular}
\negskip
\end{table}

\subsection{Varying number of steps with found LR} \label{sec:performance}

In the following experiments, we vary $n\_{steps}$ to 25\,k, 50\,k, 100\,k, and 400\,k for each network and dataset combination. Additionally, we run each experiment with 3 different random seeds, and we use the optimal LR found in the learning rate search. The networks are evaluated on the \textit{vox1-h} evaluation set.

The results are shown in Table \ref{tab:performance}. First, we see that ECAPA-TDNN has the best performance on \textit{vox2}, while wav2vec2 is slightly worse than ECAPA-TDNN but better than the X-vector network. We observe the same on the \textit{tiny-few-sessions} dataset, although the EER values are much higher compared to the other three datasets. On \textit{tiny-few-speakers} and \textit{tiny-many-sessions}, we see that wav2vec2 has the best performance, while ECAPA-TDNN is slightly worse, but better than the X-vector network. Moreover, we see that on all three tiny datasets the wav2vec2 network achieved the best performance with 50\,k steps, while ECAPA-TDNN and X-vector almost always have the best performance after 400\,k steps.


\begin{table}[]
\centering
\caption{Each network and dataset combination is trained for 25\,k, 50\,k, 100\,k and 400\,k steps with the learning rate from Table~\ref{tab:bestgrid}. The EER values are measured on the \emph{vox1-h} evaluation set. Each experiment was run 3 times.}
\label{tab:performance}
\begin{tabular}{lrrcrrcrr}
\hline
                  & \multicolumn{8}{c}{EER (mean, std in \%) on vox1-hard}\\ 
\cline{2-9}
steps  & \multicolumn{2}{c}{X-vector}                &  & \multicolumn{2}{c}{ECAPA}                   &  & \multicolumn{2}{c}{wav2vec2}\\ 
\hline 
\rlap{\textbf{vox2}}      & \multicolumn{1}{c}{} & \multicolumn{1}{c}{} &  & \multicolumn{1}{c}{} & \multicolumn{1}{c}{} &  & \multicolumn{1}{c}{} & \multicolumn{1}{c}{} \\
25k                & 16.30              & 0.64               &  & 6.80               & 0.06               &  & 7.76               & 0.07               \\
50k                & 11.21              & 0.33               &  & 5.46               & 0.07               &  & 4.66               & 0.15               \\
100k               & 7.21               & 0.10               &  & 4.61               & 0.13               &  & \textbf{4.20}      & 0.16               \\
400k               & \textbf{5.01}      & 0.04               &  & \textbf{3.93}      & 0.07               &  & 5.90               & 0.77               \\ \hline
\rlap{\textbf{tiny-few-speakers}}  & \multicolumn{1}{l}{} & \multicolumn{1}{l}{} &  & \multicolumn{1}{l}{} & \multicolumn{1}{l}{} &  & \multicolumn{1}{l}{} & \multicolumn{1}{l}{} \\
25k                & 18.93              & 0.15               &  & 18.27              & 0.03               &  & 22.48              & 0.16               \\
50k                & 18.02              & 0.08               &  & 17.59              & 0.40               &  & \textbf{15.19}     & 0.24               \\
100k               & \textbf{18.00}     & 0.22               &  & 17.48              & 0.07               &  & 15.60              & 0.20               \\
400k               & 18.04              & 0.31               &  & \textbf{16.95}     & 0.15               &  & 18.50              & 0.43               \\ \hline
\rlap{\textbf{tiny-few-sessions}}  & \multicolumn{1}{l}{} & \multicolumn{1}{l}{} &  & \multicolumn{1}{l}{} & \multicolumn{1}{l}{} &  & \multicolumn{1}{l}{} & \multicolumn{1}{l}{} \\
25k                & 28.46              & 0.12               &  & 22.67              & 0.23               &  & 23.78              & 0.10               \\
50k                & 27.23              & 0.14               &  & 21.25              & 0.44               &  & \textbf{21.88}     & 0.16               \\
100k               & 26.29              & 0.30               &  & 20.58              & 0.25               &  & 22.08              & 0.29               \\
400k               & \textbf{24.00}     & 0.19               &  & \textbf{19.52}     & 0.12               &  & 23.31              & 0.10               \\ \hline
\rlap{\textbf{tiny-many-sessions}} & \multicolumn{1}{l}{} & \multicolumn{1}{l}{} &  & \multicolumn{1}{l}{} & \multicolumn{1}{l}{} &  & \multicolumn{1}{l}{} & \multicolumn{1}{l}{} \\
25k                & 18.00              & 0.11               &  & 11.51              & 0.39               &  & 10.41              & 0.52               \\
50k                & 16.05              & 0.31               &  & 9.78               & 0.06               &  & \textbf{6.72}      & 0.04               \\
100k               & 13.53              & 0.73               &  & \textbf{9.12}      & 0.11               &  & 7.66               & 0.83               \\
400k               & \textbf{10.58}     & 0.25               &  & 9.36               & 0.08               &  & 7.93               & 0.37               \\ \hline
\end{tabular}
\negskip
\end{table}

\subsection{Ablation study} \label{sec:ablation}

For the last set of experiments we perform an ablation study on the baseline training protocol described in \cref{sec:trainprotocol}. We perform the ablations on the \textit{tiny-few-speakers} and \textit{tiny-many-sessions} datasets with the wav2vec2 network. All ablations use 50\,k steps, and the best LR found in the grid search. In the first set of ablations, we vary the learning rate schedule, and instead use either 1) a constant schedule, 2) an exponentially decaying schedule, or 3) a cyclic schedule with one cycle instead of four cycles. In the second set of ablations, we focus on the weights, and therefore we 1) randomly initialize the wav2vec2 network,  2) use self-supervised pre-trained weights but without any freezing, 3) use the pre-trained weights and freeze the feature extractor CNN for all 50\,k steps, or 4) use the pre-trained weights and freeze the network for the 1st learning rate cycle (12.5\,k steps). The third set of ablations consider regularisation. We 1) disable all regularisation parameters mentioned in \cref{sec:trainprotocol}, or 2) only enable dropout, 3) only enable LayerDrop, or 4) only enable masking. 

The results are shown in Table \ref{tab:ablation}. When we ablate on the learning rate schedule, we observe that for \textit{tiny-few-speakers} all three schedules perform worse than the baseline. For \textit{tiny-many-sessions}, an exponentially decaying schedule seems to perform slightly better than our baseline, while a constant schedule, as well as a cyclic schedule with 1 cycle, perform worse. Next, we looked at the network weights and the freezing schedule. We observe that using randomly initialized weights prevents convergence to a reasonable performance. When using the self-supervised pre-trained weights without any freezing, the performance is slightly worse than the baseline. Freezing the whole network for the first learning rate cycle seems beneficial, as it improves on the baseline, while freezing the feature extractor CNN for $n\_{steps}$ results in degraded performance. Finally, we observe that disabling all regularisation degrades performance. With only enabling LayerDrop regularization we achieve similar performance to the baseline, while only enabling masking, and only enabling dropout, perform similar to disabling regularisation.    

\begin{table}[t]
\centering
\caption{Ablation on the wav2vec2 network trained on the \emph{tiny-few-speakers} and \emph{tiny-many-sessions} datasets. Evaluation is done on the \emph{vox1-h} evaluation set. Each experiment is run 3 times.}
\label{tab:ablation}
\negskip
\begin{tabular}{llrrlrr}
\hline
                    &  & \multicolumn{5}{c}{EER (mean, std in \%) on \textit{vox1-h}} \\ \cline{3-7} 
ablation                &  & \multicolumn{2}{c}{few-speakers}                & \multicolumn{1}{c}{} & \multicolumn{2}{c}{many-sessions}               \\ \hline
baseline (Table \ref{tab:performance})           &  & 15.19   & 0.24   &    & 6.72    & 0.04   \\ \hline
\textbf{LR schedule}    &  & \multicolumn{1}{l}{} & \multicolumn{1}{l}{} &                      & \multicolumn{1}{l}{} & \multicolumn{1}{l}{} \\
constant            &  & 16.77   & 0.26   &    & 8.80    & 0.44   \\
exp. decay          &  & 16.68   & 0.20   &    & 6.67    & 0.04   \\
1 cycle             &  & 15.79   & 0.23   &    & 8.59    & 0.23   \\ \hline
\textbf{weights}        &  & \multicolumn{1}{l}{} & \multicolumn{1}{l}{} &                      & \multicolumn{1}{l}{} & \multicolumn{1}{l}{} \\
random init         &  & 33.35   & 0.16   &    & 46.24   & 0.06   \\
pretraining BASE    &  & 15.16   & 0.17   &    & 7.18    & 0.19   \\
\quad and freeze CNN       &  & 15.48   & 0.25   &    & 7.83    & 0.45   \\
\quad or freeze 1st cycle &  & 14.52   & 0.11   &    & 6.38    & 0.17   \\ \hline
\textbf{regularisation} &  & \multicolumn{1}{l}{} & \multicolumn{1}{l}{} &                      & \multicolumn{1}{l}{} & \multicolumn{1}{l}{} \\
none                &  & 16.67   & 0.27   &    & 7.73    & 0.07   \\
dropout             &  & 16.67   & 0.11   &    & 8.01    & 0.12   \\
layerdrop           &  & 15.03   & 0.14   &    & 6.72    & 0.21   \\
masking             &  & 16.21   & 0.18   &    & 7.87    & 0.24   \\ \hline
\end{tabular}
\negskip
\end{table}

\section{Conclusion}

Similar to ASR~\cite{Baevski2020Wav2vec2}, we have shown that the wav2vec2 network, when initialised with self-supervised weights, has better performance, and needs fewer training iterations, than the X-vector and ECAPA-TDNN network on two out of the three tiny datasets. 
Although ECAPA-TDNN performed slightly better on the \textit{tiny-few-sessions} dataset, the performance of all three networks was poor.
As indicated by the results on \textit{tiny-few-sessions}, a dataset with many speakers but no session variability leads to poor performance. 
As all networks had better performance on \textit{tiny-few-speakers} compared to \textit{tiny-few-sessions}, it seems that having more sessions in a limited dataset should be a priority above having many speakers.
However, \textit{tiny-few-speakers} has a mean of 51 sessions per speaker, compared to a mean of 8 for \textit{tiny-many-sessions}, while achieving worse EERs. 
It would be interesting if future work can find an optimal amount of sessions per speaker.

Currently, the self-supervised learning optimization of wav2vec2 uses a contrastive loss to distinguish a masked segment from other segments in the same utterance. For speaker recognition, it might be beneficial to include segments from other utterances, which could model inter- and intra-speaker variance. Such a change could then perhaps result in even better performance on the tiny datasets, and specifically on \textit{tiny-few-sessions}. We are also interested in future work pre-training on a dataset other than LibriSpeech, which has limited variability per speaker. It might also be relevant to pre-train on VoxCeleb2, so that the model has a prior on speech patterns, requiring less fine-tuning.

\bibliographystyle{IEEEtran}

\parskip = 0pt
\bibliography{main}


\end{document}